\def\PRD{{\em Phys. Rev.} D}

\def\NPB{{\em Nucl. Phys.} B}

\def\JHP{{\em JHEP} }
\def\CTP{{\em Commu.Theor.Phys} }

\def\PRC{{\em Phys. Rev.} C}
\def\PRL{{\em Phys. Rev. Lett.} }
\def\NPPS{{\em Nucl.Phys.Proc.Suppl} }
\def\ZP{{\em Z.Phys} }
\def\IJMP{{\em Int. J. Mod. Phys} }

\def\etal{{\it et al.}}
\documentclass[twocolumn,twoside,slac_two]{revtex4}
\usepackage{graphicx}
\usepackage{fancyhdr}
\renewcommand{\headrulewidth}{0pt}
\renewcommand{\footrulewidth}{0pt}

\usepackage{epsfig}

\input{epsf}
\begin{document}

\title{New approach in
the extracting of parton densities, based on the parameterized
solution of inverse Mellin technique}

\author{A. Mirjalili\footnote{A.Mirjalili@Yazduni.ac.ir}}
\affiliation{Physics Department, Yazd University 89195-741, Yazd,
Iran \& \\ Institute for Studies in Theoretical Physics and
Mathematics (IPM)\\\\P.O.Box 19395-5531, Tehran, Iran }
\author{K. Keshavarzian}
\affiliation{Physics Department, Azad University, Yazd Unit, Yazd,
Iran}
\begin{abstract}
We analysis  the sea quark densities, based on the  constituent
quark model. To perform a direct fit with available experimental
data, the parameterized Inverse Mellin  technique is used. The
calculation is extended  to the NLO approximation for the singlet
and non-singlet cases in DIS phenomena. We employ the approach of
complete RG improvement(CORGI) where one is forced to identify and
resum to all-orders RG-predictable ultraviolet logarithm terms
which truly build the Q-dependence of QCD observable. The results
are compared with the standard approach of perturbative QCD in the
${\overline{MS}}$ scheme with a physical choice of scale. The
results in the CORGI approach indicate a better agreement to the
data.
\end{abstract}

\maketitle

\thispagestyle{fancy}


\section{Introduction}
One of the feature of strong interaction field which is
investigated by many phenomenological model is the Quark parton
model. To  construct the hadron structure from the parton
densities, we use from  the constituent quark model \cite{Hwa-81}
which seems to give us a better insight. A constituent parton is
defined as a cluster of valence quarks accompanied by a cloud of
sea quarks and gluons. They have been referred to as valons. It
can be considered as a bound state in which, for instance, a
proton consists of three valons, two U-valons and one D-valon
which, on the one hand, interact with each other in a way that is
characterized by the valon wave function and which on the other
hand contribute independently in an inclusive hard collision with
a $Q^2$ dependence that can be calculated in QCD at high $Q^2$.
These valons thus carry the quantum numbers of the respective
valence quarks. Hwa \cite{Hwa-80} found evidence for the valons in
the deep inelastic neutrino scattering data, suggested their
existence and applied it to a variety of phenomena. Hwa
\cite{Hwa-80-II} has also successfully formulated a treatment of
the low-$P_T$ reactions based on a structural analysis of the
valons. Some papers can be found in which the valon model has been
used to extract new information for parton distributions and
hadron structure functions in unpolarized and
polarized cases\cite{Arash-I}.\\

To improve and increase the reliability of the perturbative QCD
calculation, we try to use the Complete Renormalization Group
Improvement (CORGI) approach which is related to  the long
standing problem of renormalization ($\mu$)and factorization (M)
scale dependence in QCD predictions. Usually in Renormalization
Group(RG)-improving perturbation theory, it is assumed that these
scales are related to the physical scale Q and ${\mu}$= M=Q is
normally chosen. The resulting fixed-order predictions depend on
the choice of scales. If instead one insists that these
dimentionful scales are independent of Q, one is forced to
identify and resum to all-orders RG-predictable ultraviolet
logarithms of Q which truly build the Q-dependence. In so doing
all dependence on ${\mu}$ and M disappear. This Complete RG
Improvement (CORGI) approach has previously been applied to the
single scale case where there is only ${\mu}$-dependence
\cite{Max-98}. It is then extended  to the more complicated
pattern of logarithms involved in the two scales problem of
moments of structure functions \cite{max-mir-00}. In this approach
the standard perturbative series of QCD observable is
reconstructed in terms of scheme-invariant quantities. So it is
expected to get more accurate results compared to those obtained
in the standard perturbative QCD approach.\\

Finally, in order to obtain directly all unknown parameters of the
model, just by using the available experimental data, we use from
the Inverse Mellin Technique, not in a numerical form but in a
parameterized form. Usually one uses numerical computation to
extract parton densities from their moments, but we believe that
the parameterized Inverse Mellin technique is a mathematically
powerful tool which produces more reliable results than those
obtained from the customary numerical method. Using the symmetry
properties of the inverse Mellin transformation and also a Taylor
expansion of the integrand function, we are able to obtain the
parameterized solution.
\section{Phenomenological constituent quark model}
The idea of quark cluster is not new. In this model, the hadron is
envisaged as a bound state of valence quark clusters. For example
the bound state of $\pi^-$ consists of a``anti-up" and ``down"
constituent quarks. In the static problems there is a little
difference between the usual constituent quarks and the valons,
since the point-like nature of the constituent quarks is not a
crucial aspect of the description, and has been assumed mainly for
simplicity. But, in the scattering problems it is important to
recognize that the valons, being clusters of partons, can not
easily undergo scattering as a whole. The fact that the
bound-state problem of the nucleon can be well described by three
constituent quarks implies that the spatial extensions of the
valons do not overlap appreciably. A physical picture of the
nucleon in terms of three valons is then quite analogous to the
usual picture of the deuteron in terms of two nucleons.\\

To facilitated the phenomenological analysis the following  simple
form for the exclusive constituent quark inside the proton is
assumed \cite{Hwa-81}
\begin{equation}
 {G_{UUD/p}(y_1, y_2, y_3)=g \, (y_1 y_2)^{\alpha} \, y_3^{\beta}
\, \delta (y_1+y_2+y_3-1),}\nonumber
\end{equation}
where $\alpha$ and $\beta$ are two free parameters and $y_i$ is
the momentum fraction of the i'th constituent quark. The $U$ and
$D$ type inclusive constituent quark distributions can be obtained
by double integration over the specified variables:

\begin{eqnarray}
G_{U/p}(y)&&=\int dy_2\, \int dy_3\, G_{UUD/p}(y, y_2,
y_3)=\nonumber\\ &&gB(\alpha +1,\beta +1)\, y^{\alpha}\,
(1-y)^{\alpha +\beta +1},
\end{eqnarray}
\begin{eqnarray}
G_{D/p}(y)&&=\int dy_1\, \int dy_2\, G_{UUD/p}(y_1, y_2,
y)=\nonumber\\&&gB(\alpha +1,\alpha +\beta +2)\, y^{\beta}\,
(1-y)^{2\alpha +1}.
\end{eqnarray}
The normalization parameter $g$ has been fixed by requiring
\begin{equation}
\int_0^1 G_{U/p}(y) \, dy=\int_0^1 G_{D/p}(y) \, dy=1,
\end{equation}
and is equal to $g=[B(\alpha +1, \beta +1) \, B(\alpha +1, \alpha
+\beta +2)]^{-1}$, where $B(m, n)$ is the Euler-beta function. Hwa
and Yang \cite{Hwa-02} have recalculated the unpolarized valon
distribution inside the proton with minimization method, using
direct fit to CTEQ parton distributions, and have reported a new
values of $\alpha$ and $\beta$ which are found to be $\alpha
=1.76$ and $\beta =1.05$.\\

This model  suggests that the structure function of a hadron
involves a convolution of two distributions. Constituent quark
distributions in the proton and structure function for each
constituent quark so as
\begin{equation}
F_2^p(x,Q^2)=\sum_v \int_x^1 dy \, G_{v/p}(y) \, F_2^v(z={x\over
y},Q^2)\, . \label{conv-int}
\end{equation}
We shall also assume that the three valons carry all the momentum
of the proton. This assumption is reasonable provided that the
exchange of very soft gluons is responsible for the binding.
Eq.~(\ref{conv-int}) involves also the assumption that in the deep
inelastic scattering at high $Q^2$ the valons are independently
probed, since the shortness of interaction time makes it
reasonable to ignore the response of the spectator valons. Thus,
through Eq.~(\ref{conv-int}) we have broken up the hadron
structure problem into two parts. One part represented by
$G_{v/p}(y)$, describes the wave functions of the proton in the
valon representation. It contains all the hadronic complications
due to the confinement. It is independent of $Q^2$ or the probe.
The other part represented by $F_2^v(z={x\over y},Q^2)$, describes
the virtual QCD processes of the gluon emissions and quark-pair
creation. It refers to an individual valon independent of the
other valons in the proton and consequently also independent of
the confinement problem. It depends on $Q^2$ and the nature of the
probe.\\ 

Since the calculation in moment n-space is easier than the
calculation in x-space, we work with to the moments of the
distribution, defining
\begin{equation}
M_{2,3}(n,Q^2)=\int_0^1 dx \, x^{n-2} \, \left \{
\begin{array}{c} F_2 \\ xF_3\end{array} \right \} (x, Q^2),
\nonumber
\end{equation}
It then follows from Eq.~{(\ref{conv-int})} that
\begin{equation}
M^N(n,Q^2)=\sum_v M_{v/N}(n) \, M^v(n, Q^2). \label{mom-conv}
\end{equation}
The distributions we shall calculate (all referring to the proton)
are those for sea quarks
$xu_{sea}(x)=x\bar{u}(x)=xd_{sea}=x\bar{d}(x)=xs(x)=x\bar{s}(x)=\cdots$,
which we shall generally denote by $x\bar{q}(x)$. Its moment is
denoted by $M_{sea}(n,Q^2)$ or equivalently $M_{sea}(n,s)$ where
$s$ is evolution parameter which in leading order is defined by
\begin{equation}
s=\log{{\log{Q^2/\Lambda_{\overline{MS}}^2}}\over{\log{Q_0^2/\Lambda_{\overline{MS}}^2}}}
\, ,
\end{equation}
and generally at any required order is defined by
\begin{equation}
s=\log{{a(Q_0^2)}\over{a(Q^2)}} \label{evol} \, ,
\end{equation}
where $a(Q^2)$ is related to the strong coupling constant by
$a(Q^2)=\alpha_s(Q^2)/4\pi$ . Using  Eq.~{(\ref{mom-conv})}, we
obtain
\begin{eqnarray}
M_{sea}(n,s)&=&{1 \over 2f} \left [2U(n)+D(n) \right ] \nonumber\\
&&\times\left [M^S(n,s)-M^{NS}(n,s) \right ],\label{mom-sea}
\end{eqnarray}
where $M^S$ and $M^{NS}$ are singlet and non-singlet evolution
function given in leading order by
\begin{equation}
M^{NS}(n, s)=\exp{(-d_{NS} \, s)},\nonumber
\end{equation}
\begin{equation}
M^{S}(n, s)={1 \over 2}(1+\rho) \, \exp{(-d_+ \, s)}+{1 \over
2}(1-\rho) \, \exp{(-d_- \, s)}.\nonumber
\end{equation}\\

These moments are the leading order solution of the
renormalization group equation in QCD. The anomalous dimensions,
d± and other associated parameters are defined in \cite{Hwa-80-II}.\\

We shall from now indicate the moment of sea quark densities by
$M(n,Q^2)$. In the NLO approximation, we have\cite{mom-NLO}\\
\begin{equation}
\begin{array}{rl}
M^{NS}(n,Q^2)=&\left[1+{\alpha_s(Q^2)-\alpha_s(Q_0^2)\over
4\pi}\left ({\gamma_1^{NS}\over 2\beta_0}-{\beta_1
\gamma_0^{qq}\over 2\beta_0^2} \right ) \right ]\\\\&\left
({\alpha_s(Q^2)\over \alpha_s(Q_0^2)} \right
)^{\gamma_0^{qq}/2\beta_0},\nonumber
\end{array}\label{NLO-mns}
\end{equation}
and\\
\begin{equation}
\begin{array}{rl}
\displaystyle \begin{pmatrix} {M^S(n,Q^2) \cr M^g(n,Q^2)}
\end{pmatrix}&\hspace{-0.25 cm}= \displaystyle \left[P_- -{1\over
2\beta_0}{\alpha_s(Q_0^2)-\alpha_s(Q^2)\over
4\pi}P_-.\gamma .P_-\right.\vspace*{5mm} \\
&\hspace{-1.5 cm}\displaystyle -\left({\alpha_s(Q_0^2)\over
4\pi}-{\alpha_s(Q^2)\over 4\pi}\left({\alpha_s(Q^2)\over
\alpha_s(Q_0^2)}\right)^{(\lambda_+ -\lambda_-)/2\beta_0}
\right)\vspace*{5mm} \\
&\hspace{-1.75 cm}\displaystyle  \left. .{P_-.\gamma .P_+\over
2\beta_0 +\lambda_+ -\lambda_-}
\right]\left({\alpha_s(Q^2)\over\alpha_s(Q_0^2)}
\right)^{\lambda_- /2\beta_0}+(+\longleftrightarrow -).
\end{array}\label{NLO-ms}
\end{equation}
The parameters ($\gamma$'s, $\lambda$'s, $\beta$'s and $P$'s) are
defined in references
\cite{mom-NLO,gam-e+e-,ver-01,ver-04-ns,ver-04-s}. \\

We will discuss how we can extend and improve the precision of the
calculation by direct use of available experimental data
\cite{mishra}, without referring to the input scales $Q_0$ and
$\Lambda_{\overline{MS}}$. In this sense, we extract parton
densities from the analytical moments of densities. A method
devised to deal with this situation is to perform the integration
of the inverse Mellin transformation in parameterized form which
will be explained in section 4.
\section{Renormalization and factorization scale  dependence}
The problem of renormalization scheme dependence in QCD
perturbation theory remains on obstacle to making precise tests of
the theory. It was pointed out \cite{Max-98} that the
renormalization scale dependence of dimensionless physical QCD
observables, depending on a single energy scale $Q$, can be
avoided provided that all ultraviolet logarithms which build the
physical energy dependence on $Q$ are resummed. This was termed
complete Renormalization Group (RG)-improvement approach. 
For a single case of a dimensional observable ${{\cal{R}}(Q)}$
with
\begin{equation}
{{\cal{R}}(Q)}=a+{r_1}{a^2}+{r_2}{a^3}+\ldots+{r_n}{a^{n+1}}
+\ldots\;.\label{obs-snd}
\end{equation}
The RS can be labelled by the non-universal coefficients of the
beta-function and $\Lambda_{RS}$ \cite{Stevenson}.
Self-consistency of perturbation theory that is the derivative of
N-th order approximant ${{\cal{R}}(Q)}$ with respect to the scheme
labelled parameters, is of higher order than the approximant
itself, will yields partial differential equations for
coefficients $r_2$, $r_3$ and ... with respect to non-universal
coefficients of $\beta$-function and $r_1$ coefficient. On
integration of these partial differential equations, one finds
\cite{max-mir-00}
\begin{eqnarray}
{r_2}({r_1},{c_2})={r_1}^{2}+c{r_1}+{X_2}-{c_2}\nonumber
\end{eqnarray}
\begin{eqnarray}
{r_3}({r_1},{c_2},{c_3})&=&{r_1}^{3}+{5\over2}c{r_1}^{2}
+(3{X_2}-2{c_2}){r_1}+\nonumber\\
&&{X_3}-{1\over2}{c_3}\nonumber
\end{eqnarray}
\begin{eqnarray}
\hspace{-2.5 cm}\hspace{5.5 cm}\vdots\;\;.
\end{eqnarray}
General Structure is as follows
\begin{eqnarray}
{r_n}({r_1},{c_2},{\ldots},{c_n})&=&{{\hat{r}}_{n}}({r_1},{c_2},
{\ldots},{c_{n-1}})+{X_n}
\nonumber\\
&&-{c_n}/(n-1)\nonumber;
\end{eqnarray}
where $X_n$ are Q-independent and RS-invariant and are unknown
unless a complete  $N^{n}LO$ calculation has been performed. Now
we can reformulated $R(Q)$ as it follows \begin{equation}
{\cal{R}}(Q)=a+{r_1}{a^2}+({r}_{1}^{2}+c{r_1}+{X_2}-{c_2}){a^3}
+{\ldots}\;. \label{R-subs} \end{equation} Given a NLO
calculation, ${r_1}$ is known but ${X_2},{X_3},{\ldots}$ are
unknown. Thus the complete subset of known terms in
Eq.~{(\ref{R-subs})} at NLO is
\begin{eqnarray}
{a_0}{\equiv}a+{r_1}{a^2}+({r}_{1}^{2}+c{r_1}-{c_2}){a^3}
+{\ldots}\;, \end{eqnarray} and it is RS-invariant. Choose $r_1=0,
c_2=c_3=…=c_n=0$ we obtain $a_0=a$ \cite{max-mir-00}. At NNLO
calculation $X_2$ is unknown. Further infinite subset of terms are
known and can be resummed to all orders,
\begin{equation}
{X_2}{a_0}^{3}={X_2}{a^3}+3{X_2}{r_1}{a^4}+\ldots\;.
\end{equation} Finally we will arrive at \begin{equation}
{\cal{R}}(Q)={a_0}+{X_2}{a_0}^{3}+{X_3}{a_0}^{4}+{\ldots}
+{X_n}{a_0}^{n+1}+{\ldots}\;,\label{obs-cgi}\end{equation} which
in fact is the expansion of QCD observable in Complete
RG-Improvement (CORGI) approach. Here $a_0=a(0,0,0, …)$ is the
coupling in this scheme and satisfies \begin{equation}
\frac{1}{a_0}+c{\ln}\left(\frac{c{a_0}}{1+c{a_0}}\right)=b{\ln}\left(
\frac{Q}{{\Lambda}_{\cal{R}}}\right)\;. \end{equation}In fact the
solution of this transcendental equation can be written in closed
form in terms of the Lambert $W$-function \cite{lam1,lam2},
defined implicitly by $W(z)\exp(W(z))=z$, \begin{eqnarray}
{a_0}&=&-\frac{1}{c[1+W(z(Q))]}
\nonumber \\
z(Q)&{\equiv}&-\frac{1}{e}{\left(\frac{Q}{{\Lambda}_{\cal{R}}}\right)}^{-b/c}\;.
\label{lamb}\end{eqnarray} where $b$ and $c$ are the first two
universal terms of QCD $\beta$-function.\\ 

Each term in Eq.~(\ref{obs-cgi}) involves a resummation of
infinite terms at specified order. For instance, the first term in
this equation, $a_0$, is representing a  resummation over NLO
contributions and the second term, $X_2a_0^3$, a resummation over
the NNLO contributions of all  terms in Eq.(\ref{obs-snd}). The
advantage of CORGI approach is that not only each term in the
related perturbative series is scheme independent but also because
it involves a resummation of RG-predictable terms, it should yield
more accurate approximations than the standard truncation of Eq.~(\ref{obs-snd}).\\

If we intend to employ the CORGI approach to extract sea quark
densities  in the LO approximation we need just to change the
RG-coupling constant $a$ to $a_0$ which exists in definition of
evolution parameter $s$ that is defined by
$s=\log{{a(Q_0^2)}\over{a(Q^2)}}$. We should note that in the
definition of $a_0$, the $\Lambda_{\cal R}$ quantity will depend
on $r_1$ and this coefficient is observable-dependent. \\

To do the NLO calculation in the standard approach  to extract sea
quark densities, it is necessary to have anomalous dimensions and
Wilson coefficient functions for the singlet and non-singlet
sectors involved, for instance, deep inelastic of lepton-nucleon
scattering or $e^+e^-$ annihilation. According to
Eqs.~(\ref{NLO-mns},\ref{NLO-ms}) and following
\cite{mom-NLO,gam-e+e-,ver-01,ver-04-ns}, the results for singlet
and non-singlet moments can be obtained  and  the final results
for sea quark density will be
independent of the chosen observable.\\

To employ the CORGI approach in higher order, we should use the
general  form for moments \cite{max-mir-02}:
\begin{equation}
\begin{array}{l}
\displaystyle M(n,Q^2)=\vspace*{3mm}\\
\displaystyle A(n)\left( {ca_0(n)\over 1+a_0(n)}
\right)^{d(n)/b}(1+X_2(n)a_0^2(n)\nonumber\\\\
+X_3(n)a_0^3(n)+\cdots+X_k(n)a_0^k(n)+\cdots),
\end{array}
\end{equation}
where $a_0(n)$ is defined by Eq.~(\ref{lamb})  and $X_k(n)$ are
the scheme invariant constants which where introduced before. For
the case of dependence on $\mu$ and $M$ scales we
have\cite{max-mir-02}
\begin{equation}
\Lambda_{\cal R}=\Lambda_{\overline{MS}}\left({2c\over b}
\right)^{-c/b} \exp{\left({d_1(n)\over bd(n)}+{r_1(n)\over
d(n)}\right)},
\end{equation}
where $d_1(n)$ is the $\overline{MS}$ NLO anomalous dimension
coefficient, and $r_1(n)$ is computed in the $\overline{MS}$
scheme with $\mu =Q$. The $(2c/b)^{-c/b}$ factor corresponds to
the standard convention for defining $\Lambda_{\overline{MS}}$. In
the NLO approximation of the  CORGI approach , we just need to
keep the first two terms in the above series. The $X_2(n)$ is
defined by \cite{max-mir-02}
\begin{eqnarray}
X_2(n)&=&{-1\over2}-{b\over 2d(n)}r_1^2(n)-{cd_1(n)\over
2b}+{d_2(n)\over 2b}+\nonumber\\
&&{c_2d(n)\over 2b}+r_2(n).
\end{eqnarray}
The NLO CORGI invariant $X_2(n)$ can be computed from the
$\overline{MS}$ results for $r_1(n)$, $r_2(n)$, $d_1(n)$ $\&$
$d_2(n)$ \cite{{ver-01}}. We should note that in
Refs.\cite{max-mir-00, max-mir-02} the adopted conventions for
defining the  anomalous dimensions and coefficient functions are
different with respect to other references. To obtain $M(n,Q^2)$
for the singlet case, we need to diagonalize the anomalous
dimension matrix. To avoid this we restrict ourselves just to
considering the non-singlet case.\\
\section{Fitting methods- Parameterized Solution} In order to
check the validity of the valon model and also to increase its
ability to get more reliable parton densities (sea quark densities
in our case), we make use of a method which we call
$\beta$-fitting.\\

We assume the following phenomenological form for the sea quark
densities:
\begin{equation}
x\bar{q}(x, Q^2)=a \, x^b \, (1-x)^c \, (1+dx+e\sqrt{x}).
\label{dis-fun}
\end{equation}
The parameters $a$, $b$, $c$, $d$ and $e$ are generally
$Q^2$-dependent. The motivation for choosing this functional form
is that the term $x^b$ controls the low-$x$ behavior of sea
densities, and $(1-x)^c$ that at large values of $x$. The
remaining polynomial
factor accounts for the additional medium-$x$ values.\\

The results of calculation indicates that the chosen functional
form will yield a better fitting for the moment of the
distribution,
than that assumed for the sea quark distribution in Ref\cite{Hwa-80}.\\

Using the Mellin transformation
\begin{equation}
M(n,Q^2)=\int_0^1 x^{n-2} \, x\bar{q}(x, Q^2) \, dx,
\end{equation}
where $x\bar{q}(x, Q^2)$ has been defined in Eq.~(\ref{dis-fun}),
we arrive at
\begin{eqnarray}
M{(n,Q^2)}&=&a \, \Gamma (1+c) \, \left ({\Gamma (-1+b+n) \over
\Gamma (b+c+n)}\right.\nonumber\\
&&\left.+{e \, \Gamma (-{1\over 2}+b+n)\over \Gamma ({1\over
2}+b+c+n)}+{d \, \Gamma (b+n)\over \Gamma (1+b+c+n)} \right ).\nonumber\\
\label{mom-disfun}
\end{eqnarray}

The unknown parameters $a,b,c,d$ and $e$ in Eq.~(\ref{mom-disfun})
are obtained from the fitting of analytical results of moments in
Eq.~(\ref{mom-sea}) over the parameterized form of this equation
which is
 in terms of
$\Gamma$ or eventually Euler Beta functions. So we call this
method $\beta$-fitting. The quantities $M^S$ and $M^{NS}$ in
Eq.~(\ref{mom-sea}), are known analytically from QCD calculation
and moments of valon distributions, $U(n)$ and
$D(n)$, are defined in \cite{Hwa-81}.\\

Now to report directly all exist parameters, just by using the
available experimental data, we can use from the Inverse Mellin
Technique, not in a numerical form but in a parameterized form.
For this propose, we take the Inverse Mellin of moment of
distribution in complex space:
\begin{equation}
F(x,Q^2)={1\over 2\pi i} \int_{c-i\infty}^{c+i\infty} {dn\over
x^{n-1}} \, M(n,Q^2)\label{inv-mel}.
\end{equation}
\\
Since the variable $c$ should be placed on the right hand side of
all singularities and considering this point that all
singularities of moments will occur for $n$ less than $2$, so we
choose $c$ equal to $3$. The integrated interval was first
$[3-10i, 3+10i]$. If we choose the interval $[3-20i, 3+20i]$,
there will be a difference only of
order about $10^{-5}$ with respect to last interval.\\

Since the function $M(n,Q^2)$ is not a simple function, our
machine will not be able to compute Eq.~(\ref{inv-mel}). We use
from this point that this integral with respect to real axis, is
symmetric. So first we do integral for interval $[c=c+0i, c+mi]$
where the final result is twice this result.
\\

The technique which we used to integrate the $M(n,Q^2)$ in
(\ref{inv-mel}) is that we first choose a small interval, say
$[c+0i, c+2\epsilon i]$. Then we expand $M(n,Q^2)$ about
$c+\epsilon i$. If $\epsilon$ is small enough, we can keep just
the first term of this expansion. In next step, we repeat the
calculation for interval $[c+2\epsilon i, c+4\epsilon i]$ and
expand $M(n,Q^2)$ about $c+3\epsilon i$. Repeating this procedure
and adding all results, we are able to calculate
Eq.~(\ref{inv-mel}) completely in parameterized form.

\begin{figure*}[t]
\centering
\includegraphics[width=56mm]{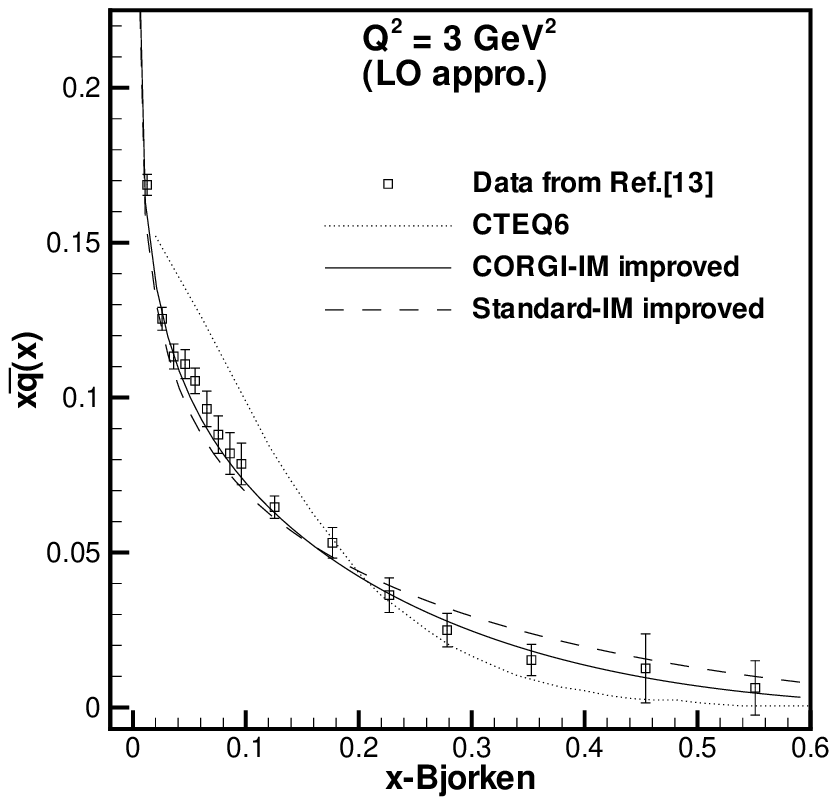}
\includegraphics[width=56mm]{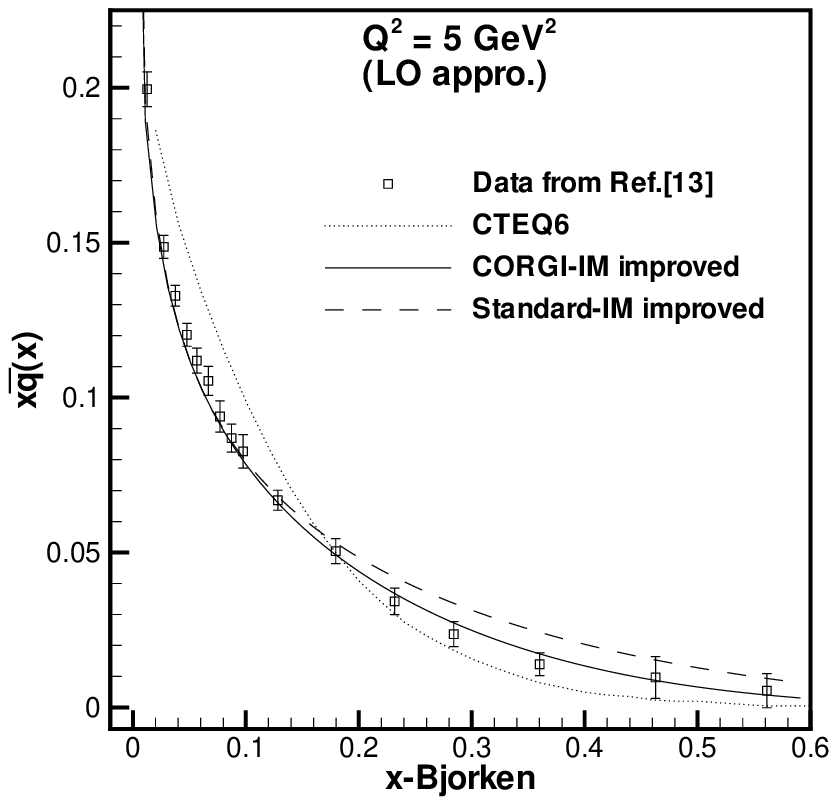}
\caption{The results of Inverse Mellin (IM) technique for the
standard and CORGI approaches in the LO approximation.}
\end{figure*}
\begin{figure*}[t]
\centering
\includegraphics[width=56mm]{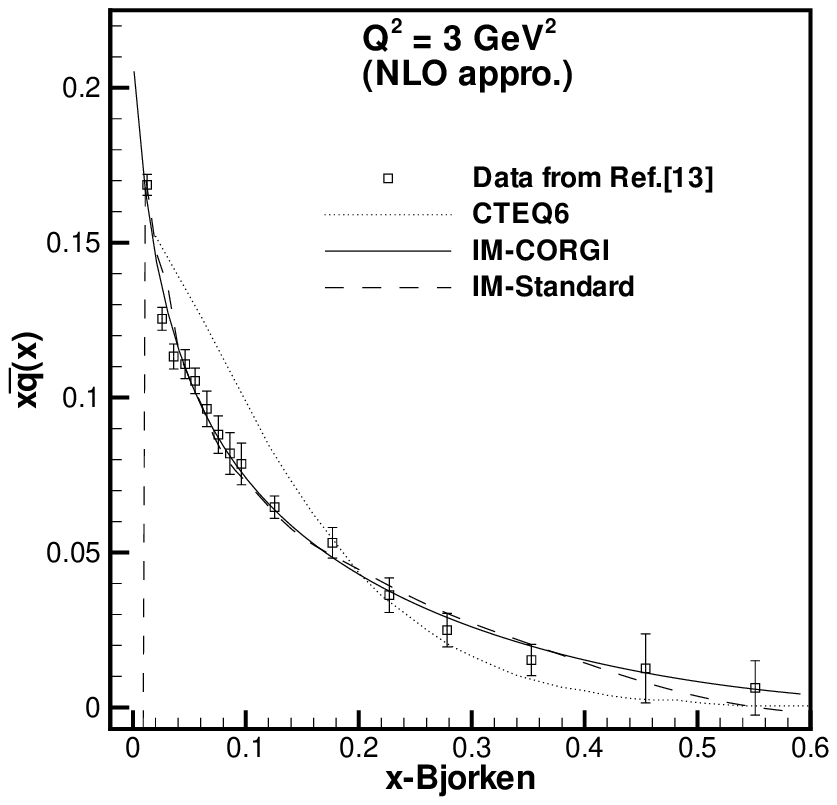}
\includegraphics[width=56mm]{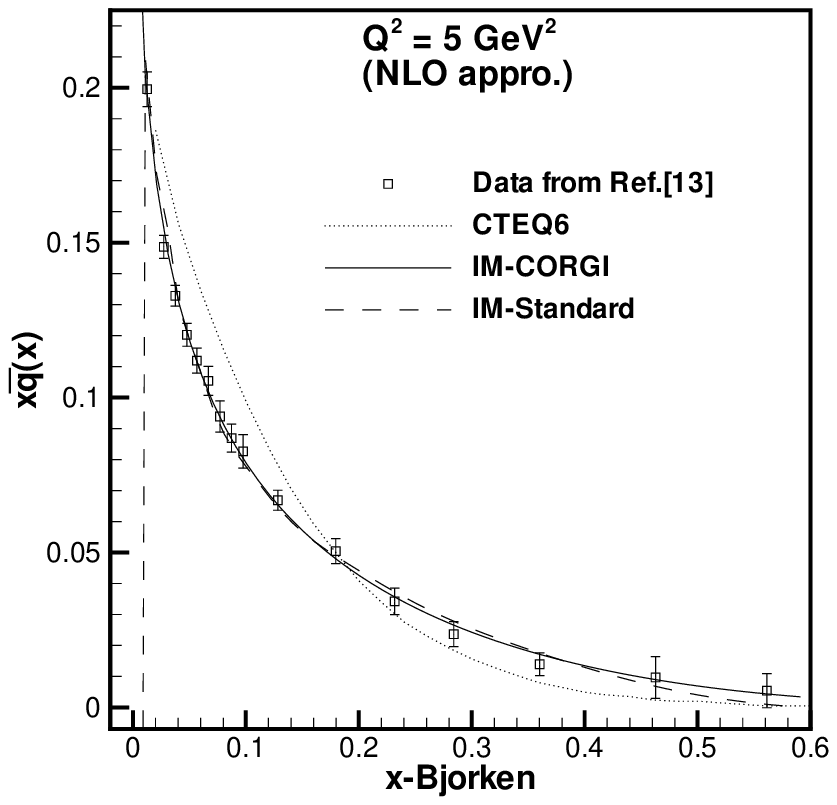}
\caption{The results of Inverse Mellin (IM) technique for NLO
approximation in both of Standard and CORGI approach.}
\end{figure*}
Now we are in a position (by direct fitting of $F(x,Q^2)$ from
Eq.~(\ref{inv-mel}) over available experimental data
\cite{mishra}) to obtain the unknown parameters of this function
which include $\alpha$ and $\beta$ (valon parameters) and $Q_0$
and $\Lambda_{\overline{MS}}$. These last two parameters occur in
the definition of the evolution parameter, $s$. Unfortunately,
from this direct fitting, we were not able to get reliable results
for the unknown parameters. To overcome this difficulty, we
inserted the values of the valon parameters $\alpha$ and $\beta$,
quoted from \cite{Hwa-02}, in the distribution function, so this
function will just depend on the $Q_0$ and
$\Lambda_{\overline{MS}}$ parameters,
$f\equiv{f(x,Q_0,\Lambda_{\overline{MS}})}$. To get the best
fitting $\chi^2$ value, we multiplied this function by an
auxiliary term, $A \, x^B \, (1-x)^C$ and fitted to obtain the
unknown parameters $Q_0$, $\Lambda_{\overline{MS}}$, $A$, $B$ and
$C$. We did the fitting for , $Q^2=3\ GeV^2$ and $Q^2=5\ GeV^2$
where we assumed a number of active quark flavours $N_f=3$ and
took an average of the fitted parameters. The difference between
the fitted parameters can be reported as an error of the
calculations. The results are tabulated in Table I.\\

One way to check the validity of calculations is to extract the
quark densities using the phenomenological form in
Eq.(\ref{dis-fun}). For this purpose, we went back to the
$\beta$-fitting method and repeated those calculations, but this
time with the  extracted values of $Q_0$ and
$\Lambda_{\overline{MS}}$ ($\alpha$ and $\beta$ taken from
Ref\cite{Hwa-02}. As a consequence, new values for parameters
contained in Eq.(\ref{dis-fun}) will be obtained. In order to get
the best fit, we again multiply Eq.(\ref{dis-fun}) by the
auxiliary term $A\, x^B\, (1-x)^C$, as we did in the inverse
Mellin Technique. We refer to this result as the ``improved
Inverse Mellin (IM) technique''. These new results in two
different standard and CORGI approaches and their comparison with
experimental data are plotted in Fig.1 and Fig.2. As can be seen
from  Fig.2 the result in CORGI approach will show better behavior
in small $x$ values.
\begin{table}[t]
\begin{center}
\caption{\it{Numerical values of the unknown parameters, resulted
from the parameterized solution of the inverse Mellein transform
technique in the LO approximation.}}\vspace{.5 cm}
\begin{tabular}{|ll|c||}
\hline  \hspace{2 cm}Fitting parameters &\\
\hline A & $3.358\pm 0.124$ \\
\hline B & $0.134\pm 0.033$ \\
\hline C & $-3.345\pm 0.264$ \\
\hline $Q_0$ & $0.422\pm 0.009$(Gev) \\
\hline $\Lambda_{\over{MS}}$ & $0.252\pm 0.004$(Gev) \\
\hline
\end{tabular}
\label{l2ea4-t1}
\end{center}
\end{table}
\section{Conclusions} Constituent quark model or valon model as
a good candidate to describe  deep inelastic scattering and to
extract sea quark densities inside the nucleon is used. The model
bridges the gap between the bound state problem and the scattering
problem for hadrons.
It is possible to use this model in the LO and NLO approximations,
using two different standard and CORGI approaches. The most
important motivation for the CORGI approach is that by completely
resumming all the UV logarithms, one correctly generates the
physical dependence of the moments ${\cal{M}}(n,{Q}^{2})$ on
the DIS energy scale $Q$.\\

In order to get the unknown parameters of the model from the
fitting over the available experimental data, parameterized
Inverse Mellin technique is used. The compared results will show
that the CORGI approach indicates better consistency with
available experimental data. An alternative ``$\beta$-fitting''
method which is based on using the defining parameters of the
Valon model, was also used to confirm the calculations of the
parameterized Inverse Mellin technique. \\

These calculations can be extended to the higher order of standard
and CORGI approaches, using the recent analytical calculations for
Wilson coefficients and anomalous dimensions.\\

\end{document}